\documentclass[seceq]{ptptex}
\usepackage{graphicx}

\title{Modified Spin-Wave Theory for Nanomagnets : Application to the Keplerate Molecule Mo$_{72}$Fe$_{30}$ }
\author{
Olivier {\sc C\'epas}\footnote{E-mail address: cepas@ill.fr}  and  Timothy {\sc Ziman}\footnote{Also at the CNRS, LPM2C, UMR 5493, Grenoble}}
\inst{Institut Laue Langevin, B.P. 156, 38042 Grenoble, France.}

\abst{
We adapt Takahashi's modified spin-wave theory to the context of nano-magnets, and apply it to the molecular compound based on the giant magnetic molecule Mo$_{72}$Fe$_{30}$. This involves solving numerically the mean-field equations and then forcing the sublattice magnetizations to zero by means of local chemical potentials for the magnons. We have thus constructed a quantum state with no local magnetization at all temperatures, appropriate to a finite-size system, but with strong correlations. We compare theoretical results to specific heat and ESR measurements.} 

\begin{document}
\maketitle

\section{Introduction}

Much effort has been devoted to low-dimensional antiferromagnetism in
the past years and difficult questions as to whether a magnetic system
of given geometry and spin-symmetry displays long-range order or
remains disordered at zero temperature were tackled. For instance, for
two dimensional isotropic Heisenberg models the triangular lattice or
the $J_1-J_2$ square model were studied by a variety of analytical and
numerical techniques in order to establish order at low
temperature\cite{MisguichLhuillier}.  These techniques include
finite-size scaling, i.e.  extrapolation from finite systems for which
exact numerical results are available to the thermodynamic limit of
physical interest.  In the newer field of nano-magnetism the question
is rather inverted: as these systems are finite, by definition there
is no long range order.  Because of thermodynamic or quantum
fluctuations it is not possible to break spontaneously any symmetry.
For large nanomolecules, there should be traces of incipient long
range order essentially in the low frequency parts of the
spectrum. The issue, then, is rather to find an accurate description
of the system and of its dynamics.

Several antiferromagnetic nano-magnets have been studied recently,
ranging from small isolated molecular units, such as V$_3$ to the
giant molecule Mo$_{72}$Fe$_{30}$ \cite{Muller}, via a large number
of intermediate-size clusters.  As for potential applications, they
are promising materials if their magnetic states can be externally
``read'', for example by using a spin-polarized current as proposed in 
multilayers.\cite{Kiselev} From a fundamental point of view, they are
interesting systems to study the crossover from simple quantum
molecular systems to larger systems that may behave as classical
N\'eel states. They can, to a first approximation, be treated
as isotropic. According to the original picture by
Anderson for systems of continuously  degenerate
broken symmetries,\cite{Anderson} a manifold of degenerate classical N\'eel states
is obtained when a family of  finite energy states (the ``tower'' of
states) of the system of size $N$ collapse onto the ground state when $N \rightarrow
+\infty$ (as $1/N$).  Systems with frustration are particularly interesting
in that  it may amplify the competition between
quantum-mechanically disordered and N\'eel-like states. There is even no guarantee that the system would order in the
thermodynamic limit.  It is also known that
competitions arise when an external magnetic field is applied to
systems such as the infinite triangular lattice \cite{Miyashita} or
the Kagom\'e lattice \cite{Zhitomirsky}. Nano-magnets are similarly
being studied under field.\cite{Nojiri} An external magnetic-field
also allows for measurements of crossing of excited states
and quantized magnetization processes.\cite{Barbara}

Mo$_{72}$Fe$_{30}$ is an abbreviation for a recently synthesized giant
molecule that consists of 30 Fe$^{3+}$ $S=5/2$ ions which occupy the
vertices of an icosidodecahedron (known as one of the Kepler solids,
hence the name Keplerate for the molecule)\cite{Muller}. Experiments
show an absence of magnetic order at low temperatures, suggesting that
the molecules of the solid are relatively isolated from one another
\cite{Muller}.  This molecule is highly frustrated, consisting of
triangle- and pentagon-sharing vertices. Thus the spins of each
nano-molecular unit may be seen as forming a finite closed surface.
When a magnetic-field is applied, there is a dip in the magnetization
as function of field at 1/3 of the saturation field, which has been
interpreted as proximity to the $\uparrow \uparrow \downarrow$
state.\cite{Nojiri} There are also predicted features such as a large
jump in the magnetization just before saturation\cite{Schnack1},
reminiscent of the divergent susceptibility in bulk
systems\cite{Zhitomirsky2}.

From the theoretical standpoint, studies assuming classical
interacting spins have been made. Axenovich and Luban\cite{Axenovich}
argued that such a lattice could sustain the same order as the
triangular lattice (sublattice spins forming 120$^{\circ}$ angles) at
zero temperature, thanks to the possibility of decomposing the
icosidodecahedron into three sublattices. By construction, such a
classical state breaks time-reversal symmetry, although this is
forbidden in a finite-size system. More recently classical Monte-Carlo
calculations have been performed at finite
temperatures.\cite{Hasegawa} As expected, the thermal fluctuations
prevent the system from ordering, but also give a large specific heat
at low temperatures. From a purely classical point of view, it is
difficult to explain simultaneously the absence of magnetic-order and
a vanishing specific heat (as observed\cite{NojiriPrivate}), because
the modes that destroy the magnetic order would contribute to the
specific heat by $k_B/2$ per mode.  Quantum fluctuations have not been
considered so far and should be able to resolve such issues. Schnack
et al. have discussed a simplified quantum model where all the spins
of a given sublattice are connected to all the spins of the other
sublattices\cite{Schnack}. Such a model is integrable and they find
rotational bands as low-lying excitations\cite{Schnack}, the first
band being confirmed by a DMRG in a more realistic Heisenberg system
with nearest neighbor interactions only\cite{Exler}. There is no
theory, however, that bridges the gap between purely classical
approaches, such as that of Ref. \cite{Axenovich,Hasegawa} and quantum
ones.

Recently, Nojiri et al. have observed optical resonances at low
temperatures.\cite{NojiriPrivate} They have interpreted these
resonances as transitions from the ground state to the first excited
state. In a purely \textit{spin-isotropic} model, such as in
Ref. \cite{Schnack}, the transition probability vanishes because of
the conservation of the total spin. To explain such transitions, one
needs to invoke the presence of anisotropic interactions, whether the
transitions be of magnetic origin as, possibly, in low-dimensional
spin-liquids\cite{Sakai} or of electric origin, which could proceed
via an anisotropic coupling with the phonons \cite{Cepas}. It seems
therefore important to consider what the possible additional couplings
to a Heisenberg Hamiltonian are and whether these corrections are
able to capture the optical processes that have been
observed. Hasegawa and Shiba have considered several anisotropic
corrections to their classical Hamiltonian \cite{Hasegawa} : the
dipole-dipole interactions could be safely neglected thanks to large Fe-Fe
distances, the first corrections being single-ion, or, possibly, of
Dzyaloshinskii-Moriya type. We have considered below the simplest
single-ion anisotropy which is allowed in $S=5/2$ systems and
often dominates the anisotropic interactions.

Such a model lacks the simplicity of that of Ref. \cite{Schnack} and
can not be treated exactly. As previously noticed, the size of the
Hilbert space, $6^{30} \sim 10^{23}$, prevents using techniques such
as exact numerical diagonalisations. We have adopted a different
approach by solving the mean-field problem (which is classical in
essence) first and introducing the quantum corrections by means of
Holstein-Primakoff bosons. Doing so, we artificially break the
symmetry by allowing the magnetic order to occur. To restore the
symmetry, we use a technique that has been introduced by
Takahashi.\cite{Takahashi0,Takahashi} It consists of enforcing
\textit{a posteriori} the magnetizations to be zero on each site.
This allows one to find phases with no sublattice magnetization,
i.e. which do not break the time-reversal symmetry.

In section \textsection \ref{meanfield}, we solve numerically the
mean-field equations for quantum spins on an icosidodecahedron and
find in particular a form of magnetic order at zero temperature which
is close to, but not exactly the same as, the 120$^{\circ}$ structure
of the triangular lattice. In section \textsection \ref{HP}, we
calculate the first quantum corrections and apply Takahashi's method
to enforce the local constraints. We discuss the excitation spectrum,
the two-point correlation functions in the ground state and some
observables such as the ESR intensity or the specific heat at
zero-field.

\section{Mean-Field Theory of the Models with Anisotropy}
\label{meanfield}
\begin{figure}[htbp]
\centerline{
  \includegraphics[width=6cm,angle=-90]{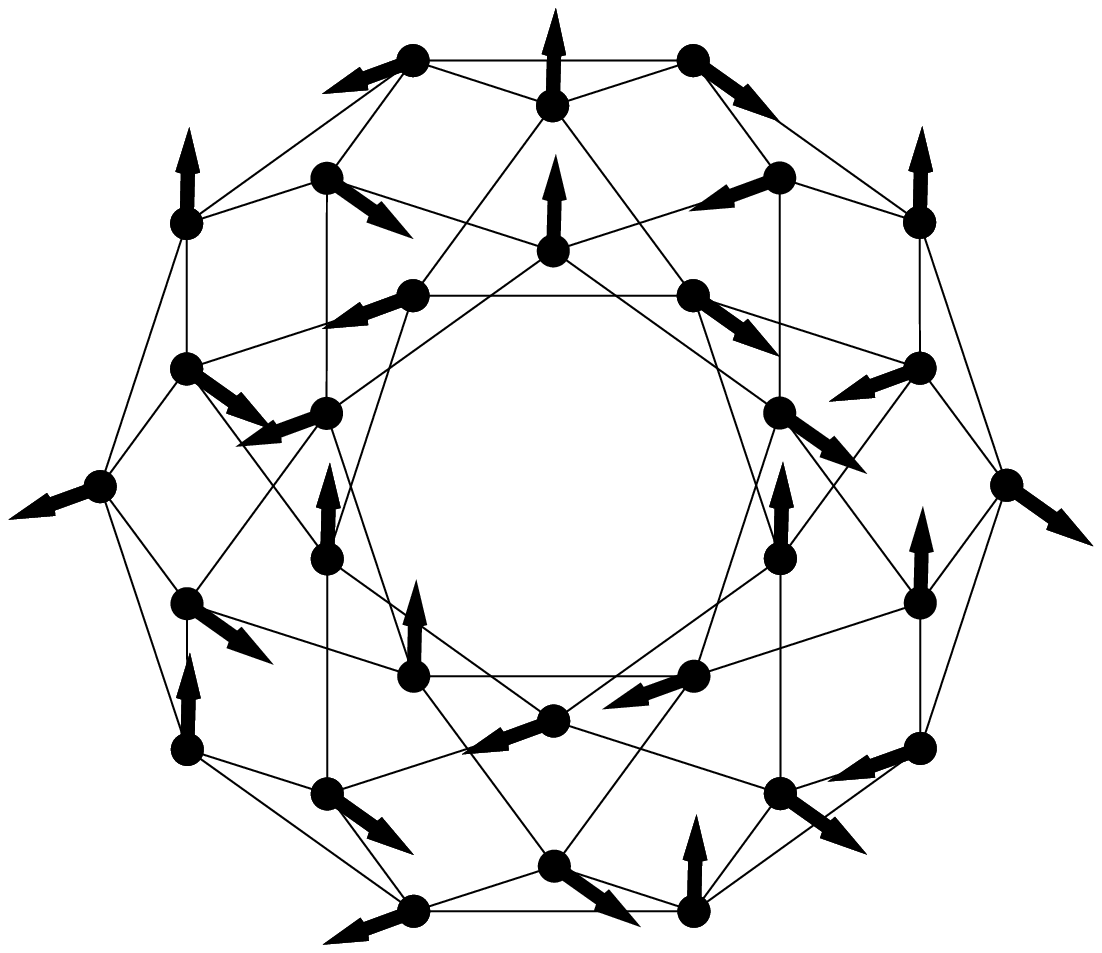}
  \includegraphics[width=6cm,angle=-90]{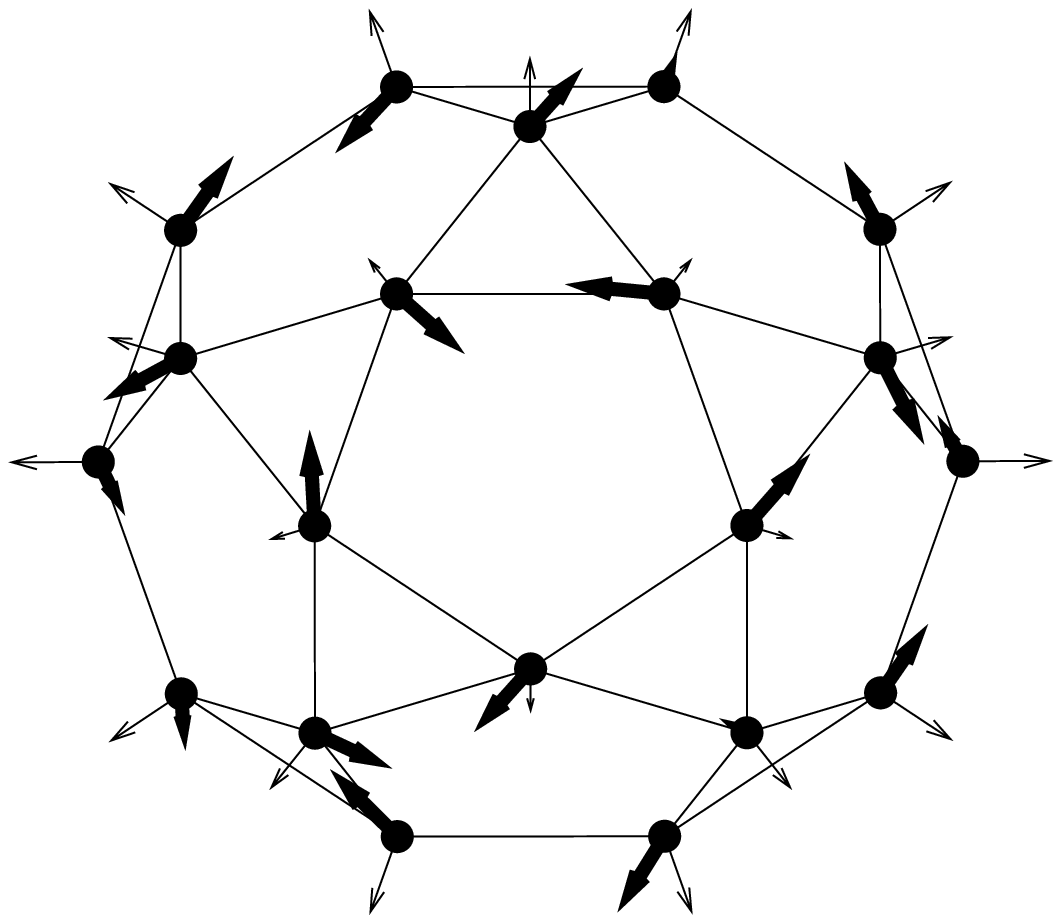}
}
\caption{Spin arrangements on the icosidodecahedron lattice (thick
vectors), calculated by mean-field theory at $T=0$ and zero external
field. Left : A \textit{global} easy-plane anisotropy identical for
all spins and perpendicular to the plane of the sheet favors a
3-sublattice \textit{coplanar} 120$^{\circ}$ state identical to that
of Ref. \cite{Axenovich}. There is a continuous degeneracy as it is
possible to rotate the spin of one sublattice in the
plane. Right: The easy-plane $D^{(i)}$ vectors (represented by thin
vectors) are \textit{local} and point towards the center of the solid
at each point. The spin order is more \textit{tangential} to the
sphere. The degeneracy is quite large.}
\label{spins}
\end{figure}

Since the molecules are relatively isolated from one another, we consider a one-molecule problem, i.e. a spin model where the spins occupy the vertices of an icosidodecahedron (Fig. \ref{spins})
\begin{equation}
H= \frac{1}{2} \sum_{<i,j>} J  \textbf{\mbox{S}}_i . \textbf{\mbox{S}}_j + \sum_{i=1}^N D^{(i)} (S_i^{(i)})^2,
\label{energy}
\end{equation}
where $\textbf{\mbox{S}}_i$ is a quantum spin operator of $S=5/2$ (of
Fe$^{3+}$), $J$ is the antiferromagnetic coupling between the nearest
neighbors $<i,j>$. The factor $1/2$ removes double counting. $N=30$ is
the number of sites. We consider two types of anisotropy :
\medskip 
\begin{enumerate}
\item $D^{(i)}$ are \textit{global} single-ion anisotropies, with the axis identical
for all sites. 
This is not very realistic given the geometry of the
icosidodecahedron, but is a simple model to compare with.
\item $D^{(i)}$ are \textit{local} single-ion anisotropies, and $(i)$
is a local direction which points towards the center of the solid at
each site, $D^i=D \hat{\textbf{\mbox{R}}}_i$, where
$\hat{\textbf{\mbox{R}}}_i$ is the position vector of site
$i$, as in Ref. \cite{Hasegawa}. 
\end{enumerate}
\medskip
In both cases, the strength of the anisotropy is small $|D|/J=0.1$,
which is justified by the fact that Fe$^{3+}$ has a closed shell
(L$_{tot}$=0). We consider easy-plane or easy-axis types.  In the
Zeeman coupling, we take $g=2$. At $T=0$, one wants to minimize the
energy (\ref{energy}). This is done by iterating numerically the $3N$
self-consistent mean-field equations for quantum spins, starting from
random initial states (up to $10^4$).

For model 1 with easy-plane anisotropy, the system converges to a
simple three-sublattice 120$^{\circ}$ magnetic order at zero temperature
(Fig. \ref{spins}, left). This is the state found in
Ref.\cite{Axenovich} except that the present anisotropy forces the
spins to lie in a plane perpendicular to the D-vector. The energy per
spin is simply $E_0/N=-S^2$. We note that replacing $S^2$ by
$S(S+1)$\cite{Axenovich,Luban} does not give a correct estimation of
the zero-point energy. For this a calculation of the quantum
corrections has to be performed (see below) and generally gives
different results.  

For model 2, the state is very close to the 120$^{\circ}$ order except
that is it more \textit{tangential} to the sphere (Fig. \ref{spins},
right).  To see it more clearly, we have depicted the distribution of
classical scalar products $ \langle
\textbf{\mbox{S}}_i. \textbf{\mbox{S}}_j\rangle= \langle
\textbf{\mbox{S}}_i \rangle . \langle \textbf{\mbox{S}}_j \rangle$ in
fig. \ref{Psisjla}. The majority of nearest neighbor bonds have a
scalar product close to $-1/2$, corresponding to an angle of
120$^{\circ}$, but some of them depart from that angle. The energy is
$E_0/N=-0.989S^2$ for $D/J=0.1$. It is interesting to note that the
state accomodates fairly well the local single-ion anisotropy: the
energy increase is a tenth of the anisotropy. We have found a large
number of degenerate states, probably owing to the large number of
symmetries of the icosidodecahedron.

\section{Holstein-Primakoff Modes in Finite-Size Systems}
\label{HP}

\subsection{Method}
\label{method}
Having solved the mean-field problem in the previous section, we can
introduce the first quantum corrections by expanding the free-energy
about a mean-field state. To do  this, we work in a new frame where the
new local coordinate axis at site $i$, called $z^{\prime}$, coincides
with the classical direction of the spin on the same site,
\cite{Walker} given by the solution of the mean-field problem. We then
use the Holstein-Primakoff representation of the spins in terms of
boson operators \textit{in the new frame}:
\begin{equation}
S^{+^{\prime}}_i = \sqrt{2S- a^{\dagger}_ia_i } \hspace{0.1cm} a_i;
\hspace{0.6cm} 
S^{-^{\prime}}_i = a^{\dagger}_i \sqrt{2S-
a^{\dagger}_i a_i}; \hspace{0.6cm} S^{z^{\prime}}_i = S -
a^{\dagger}_ia_i
\end{equation}
where the primes correspond to the new local axis. By expanding the
Hamiltonian to second-order  in the operators $a_i$,
$a_i^{\dagger}$,\footnote{Strictly speaking it is a
simplified version of Takahashi's method (see below) who, for the simpler
square lattice model, used the
Dyson-Maleev representation and took into account the interaction
between the spin-waves at the Hartree-Fock level.} it takes the form of free bosons in 
\textit{real} space:
\begin{equation}
H = E_0 + \frac{E_0}{S} + \frac{1}{2} \sum_{<i,j>} \left( A_{ij} a_i^{\dagger} a_j + A_{ij}^* a_i a_j^{\dagger} + B_{ij} a_i^{\dagger} a_j^{\dagger}  + B_{ij}^*  a_i  a_j \right) \label{hameq}  
\end{equation}
$E_0$ is the mean-field energy that we found in the previous
paragraph. The coefficients $A_{ij}$ and $B_{ij}$ (${\cal O}(S)$) are
function of the couplings and the local rotation
matrices.\cite{Walker} The Hamiltonian is Hermitian and we have
$A_{ij}=A^*_{ji}$ and $B_{ij}=B_{ji}$.  To define the fourth term of
(\ref{hameq}), we had to commute the $a_j^{\dagger}a_j$ operators. It
gives a constant term that reduces to the second term of
eq. (\ref{hameq}), $E_0/S$. Such a Hamiltonian is usually diagonalized
by a Bogoliubov transformation which ensures the bosonic characters of
the final bosons \cite{Walker,White}. Such a transformation can
generally be constructed numerically in systems with a sufficient
amount of anisotropy \cite{Watabe,Gingras}, but not directly when a
continuous symmetry leads to the existence of Goldstone modes. In this
case the bosonization procedure is singular, and it is well-known for
the infinite triangular lattice that the Hamiltonian can not be
bosonized at $k=0$ or $k=\pm 4 \pi/3$ \cite{Lhuillier,Sorella}. The
singular Goldstone modes can easily be separated out for infinite
systems, simply by omitting the $k=0,\pm4\pi/3$ modes in the sums over
$k$,\cite{Lhuillier,Sorella} but here there is no $k$-space. In any
case, in a finite-size system, the Goldstone modes are spurious since
the assumption of a broken-symmetry phase is obviously
incorrect. Takahashi \cite{Takahashi0,Takahashi} and Hirsch and Tang
\cite{Tang} have tackled this problem and imposed the condition that
the magnetization should be zero in a finite-size system,
\begin{equation}
\langle S_i^{z \prime} \rangle  = S- \langle a^{\dagger}_i a_i \rangle  =0,
\label{Neq}
\end{equation}
for all sites $i$. This is reminiscent of the paper by Rastelli and
Tassi who introduced this idea for the paramagnetic phase of a
ferromagnet \cite{Rastelli}. If in special cases, the condition could be enforced by a unique Lagrange multiplier \cite{Takahashi,Tang}, the  $N$ local
constraints usually oblige us to introduce $N$ different Lagrange multipliers $ \{ \lambda_i \}, i=1,...,N$:
\begin{equation}
H^{\prime} = H + \sum_{i=1}^N \lambda_i a^{\dagger}_i a_i 
\label{lm}
\end{equation}
The $ \{ \lambda_i \}$ can also be viewed as \textit{local}
chemical potentials for  spin-flips, since
$a_i^{\dagger} a_i$ is the number operator of spin-flips on site
$i$. To include these additional terms, the Hamiltonian in
eq. (\ref{hameq}) is modified by simply replacing $A_{ij}$ by
$A_{ij}^{\prime}=A_{ij} + \lambda_i \delta_{ij}$ and adding a constant
term to the energy, $-\frac{1}{2} \sum_i \lambda_i$.  Now all the
quantities, and the excitation spectrum in particular, will depend
upon the $\{ \lambda_i \}$.  We shall determine the set of $\{
\lambda_i \}$ by solving the $N$ equations (\ref{Neq}). To do this we
have to diagonalize the Hamiltonian $H^{\prime}$ and then calculate
the expectation values in the ground state, such as $\langle
a_i^{\dagger} a_i \rangle$.  When finite Lagrange multipliers are
added, the Goldstone modes are removed from the problem and it is
possible to construct numerically a Bogoliubov transformation. The
diagonalized Hamiltonian then takes the form:
\begin{equation}
H^{\prime} =  E_0 + \frac{E_0}{S} + \sum_{j=1}^N \omega_j \left( a_{\omega_j}^{\dagger} a_{\omega_j} + \frac{1}{2} \right) - \frac{1}{2} \sum_{j=1}^N \lambda_j
\end{equation}
where $a_{\omega_j}^{\dagger}$ is a boson operator that creates an
excitation of energy $\omega_j$. Note that the zero-point energy is
given by $E_0/S+\frac{1}{2} \sum_j (\omega_j-\lambda_j)$ (the two
terms cancel out for a simple easy-axis ferromagnet for instance). The
Bogoliubov transformation also gives the eigen-operators as function
of the local boson operators and all the expectation values of the
form $\langle a_i^{\dagger} a_j \rangle$, for instance, can be
calculated. We then solve the eqs. (\ref{Neq}) by a standard numerical
routine that finds the roots of a set of non-linear equations. Once
the Lagrange multipliers are found, the state satisfies $\langle
\textbf{\mbox{S}}_i \rangle = 0$. We then compute various physical
quantities, such as the excitation spectrum $\omega_j$, and the total
energy $E(T)$. The latter needs the explicit determination of the
Lagrange multipliers at each temperature and needs subtraction of the
magnon chemical potential part.  We also compute the two-point
correlation functions $\langle
\textbf{\mbox{S}}_i. \textbf{\mbox{S}}_j \rangle$ of the ground state.
The solution is only valid at low temperatures because the approach
starts from a low-temperature minimum that is obviously different from
the paramagnetic state. We expect the present solution to depart from
the exact solution when $T \sim T_N$ (where $T_N$ is the mean-field
N\'eel temperature of the classical system).

\subsection{Simplified Model with Global Anisotropy Axis}
\label{globalanisotropy}

We first start by giving the results of a simpler problem where all
the $D$-vectors are parallel to the same  axis (model 1). In this
case,  mean-field theory predicts a simple coplanar state with
three sublattices of spins at  120$^{\circ}$ (Fig. \ref{spins},
left). Purely classical models were similarly considered in
Refs. \cite{Axenovich,Hasegawa} and we would like to stress  what
changes 
quantum fluctuations  bring to the classical picture. First of all, the state
that we have constructed satisfies the constraint of zero
magnetization on each site, as it should  for a finite-size system.
Nevertheless, there are strong correlations that reflect features of 
the original classical state. We shall now describe these.

\begin{figure}[htbp]
\centerline{
\includegraphics[width=6cm,angle=-90]{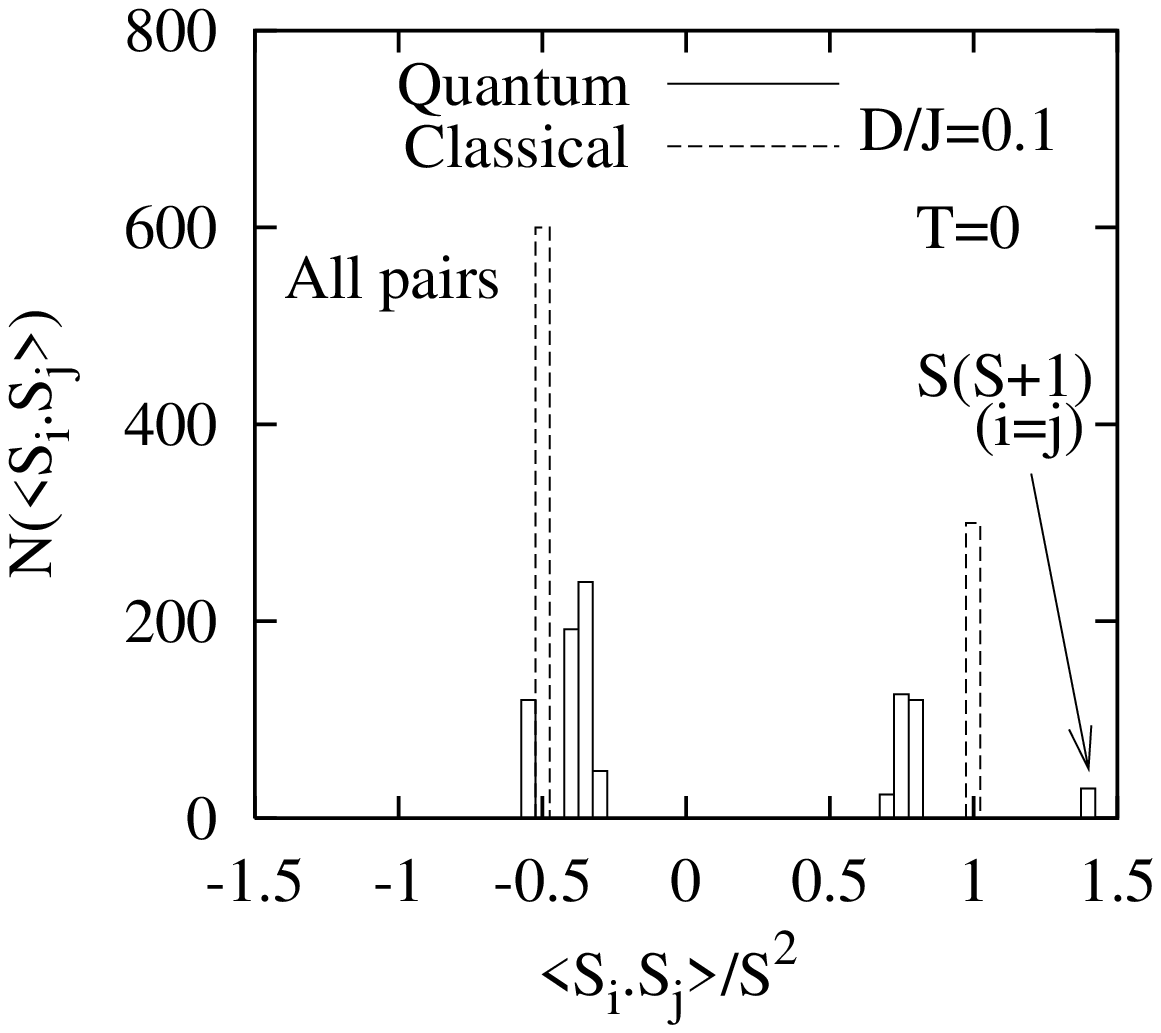}
\includegraphics[width=6cm,angle=-90]{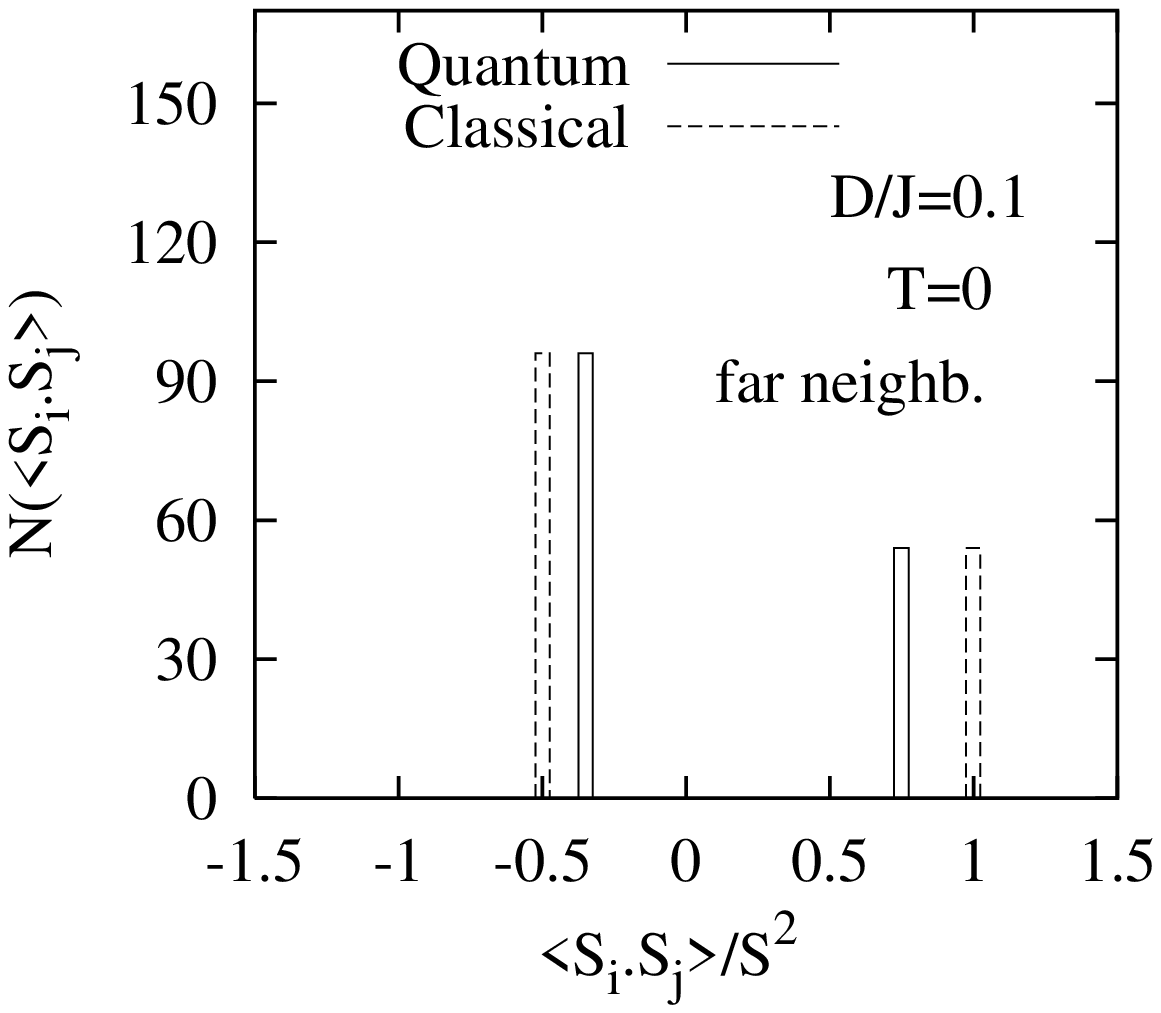}}
\caption{Number of bonds with a given value of the correlation
function for all the 900 bonds (left) and far neighbors
(right). Global anisotropy axis. The quantum correlations $ \langle
\textbf{\mbox{S}}_i. \textbf{\mbox{S}}_j\rangle$ are obtained by
applying Takahashi's method whereas the classical correlations are
simply given by $ \langle
\textbf{\mbox{S}}_i. \textbf{\mbox{S}}_j\rangle= \langle
\textbf{\mbox{S}}_i \rangle . \langle \textbf{\mbox{S}}_j \rangle$.
The classical ground state is a 3-sublattice coplanar state with
120$^{\circ}$ angles. On the right, we see that the quantum
fluctuations reduce the correlations between far neighbors, though not
to zero.}
\label{correlationsua}
\end{figure}

\textit{Correlations}. In Fig. \ref{correlationsua}, we compare the
distribution of the scalar products $\langle
\textbf{\mbox{S}}_i. \textbf{\mbox{S}}_j \rangle$ on the different
bonds for the mean-field solution and the Takahashi's method. As
expected, the quantum fluctuations have broadened the distribution
(Fig. \ref{correlationsua}, left). A question is whether or not
fluctuations have destroyed the correlations between spins that are
far away from each other. In other words, how does the correlation
length compare with the size of the molecule? In
Fig. \ref{correlationsua}, we look at pairs of spins at given
distances. The nearest neighbors are still strongly correlated, but
the correlation of far neighbors have decreased, although not to zero
(Fig. \ref{correlationsua}, right).

\begin{figure}[htbp]
\centerline{
 \includegraphics[width=6cm,angle=-90]{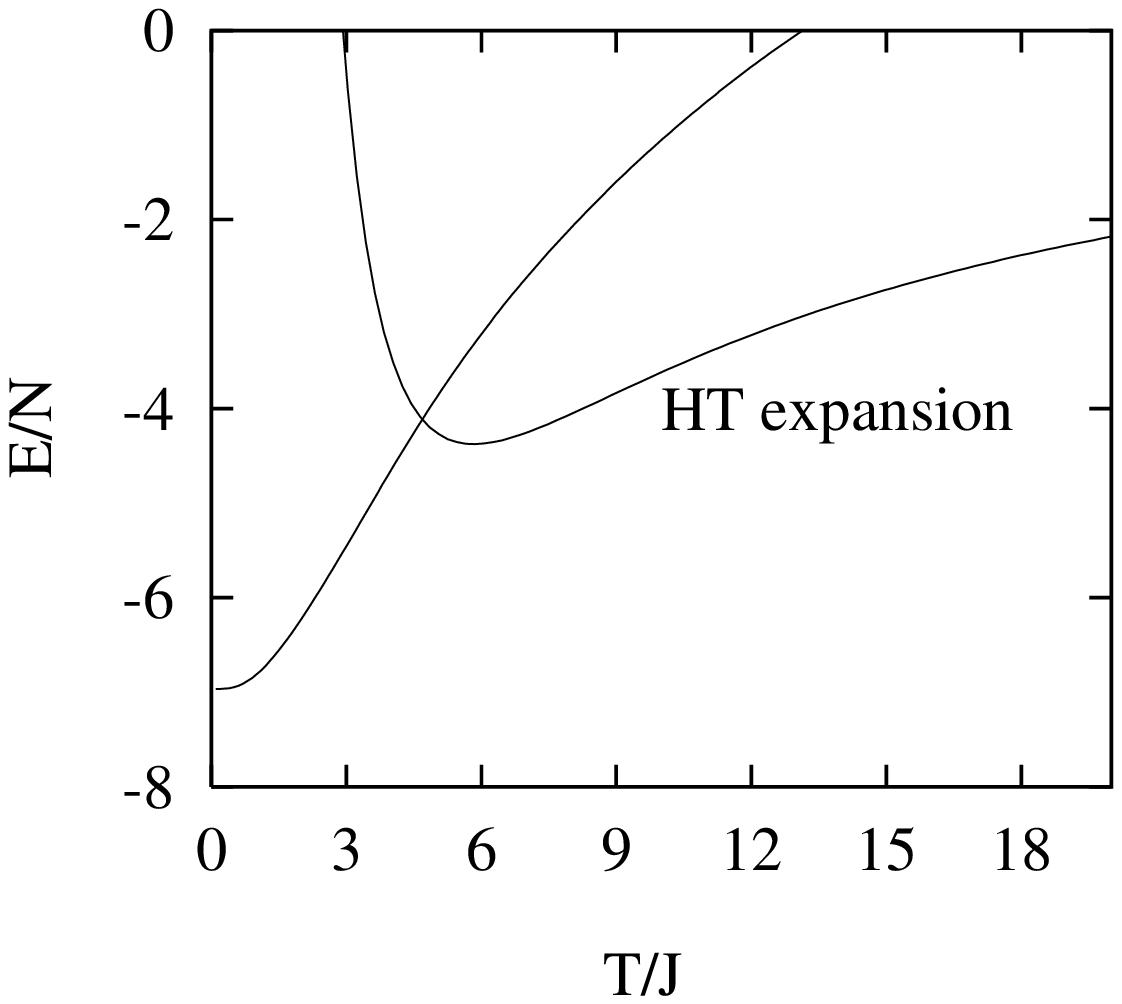}
 \includegraphics[width=6cm,angle=-90]{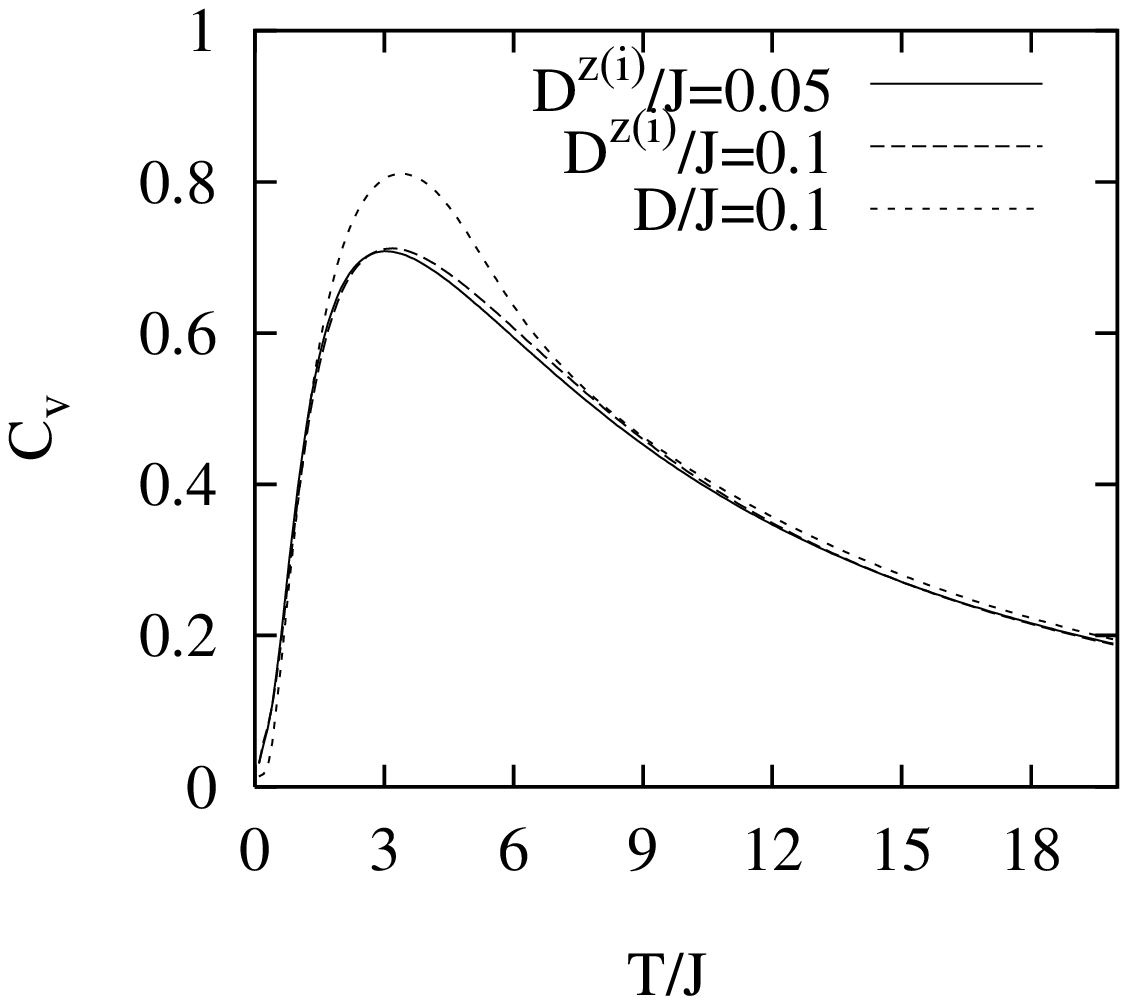}}
\caption{Energy per spin vs. temperature, calculated by modified
spin-wave theory and high-temperature (HT) expansion (left). There is
a critical temperature above which the modified spin-wave theory
breaks down. Specific heat vs. temperature for the system with global
anisotropy, $D/J=0.1$ (section \ref{globalanisotropy}) and local
anisotropy, $D^i/J=0.05$ and $D^i/J=0.1$ (section
\ref{localanisotropy}). The present theory breaks down above $T
\gtrsim 3J$ where the specific heat is overestimated. The last two
curves are almost identical, thus indicating that $C_v$ is insensitive
to the strength of the anisotropy. }
\label{specific-heat}
\end{figure}

\textit{Specific Heat}. The classical specific heat, computed by
Monte-Carlo simulation, remains finite at zero temperature and shows a
peak at $0.3J$.\cite{Hasegawa} The specific heat calculated with
quantum corrections exhibits different features
(Fig. \ref{specific-heat}, right). It vanishes at zero temperature and
the peak (which is fairly independent of the strength of the
anisotropy) is pushed to higher temperatures $\sim 3J$ , that is more
consistent with experiment. We now compare the results of the modified
spin-wave theory with that of a high-temperature expansion that is
given at second order by the analytic expression, $E/NJ = - 51.04 J/T
+ 148.87 (J/T)^2$. Note that it would be highly desirable to have
higher-order terms, that can be obtained systematically.\cite{Domb} We
plot the energies in Fig.  \ref{specific-heat}, left. We see that
there is clearly a regime at high temperature where the modified
spin-wave theory breaks down. In this regime the specific heat is
greatly over-estimated, and consequently the position of the peak may
be over-estimated too. Nonetheless, at low temperatures, the quantum
corrections should yield the correct behavior of the specific heat and
provide a reliable way of comparing with experiments.

\begin{figure}[htbp]
\centerline{
\includegraphics[width=5.5cm,angle=-90]{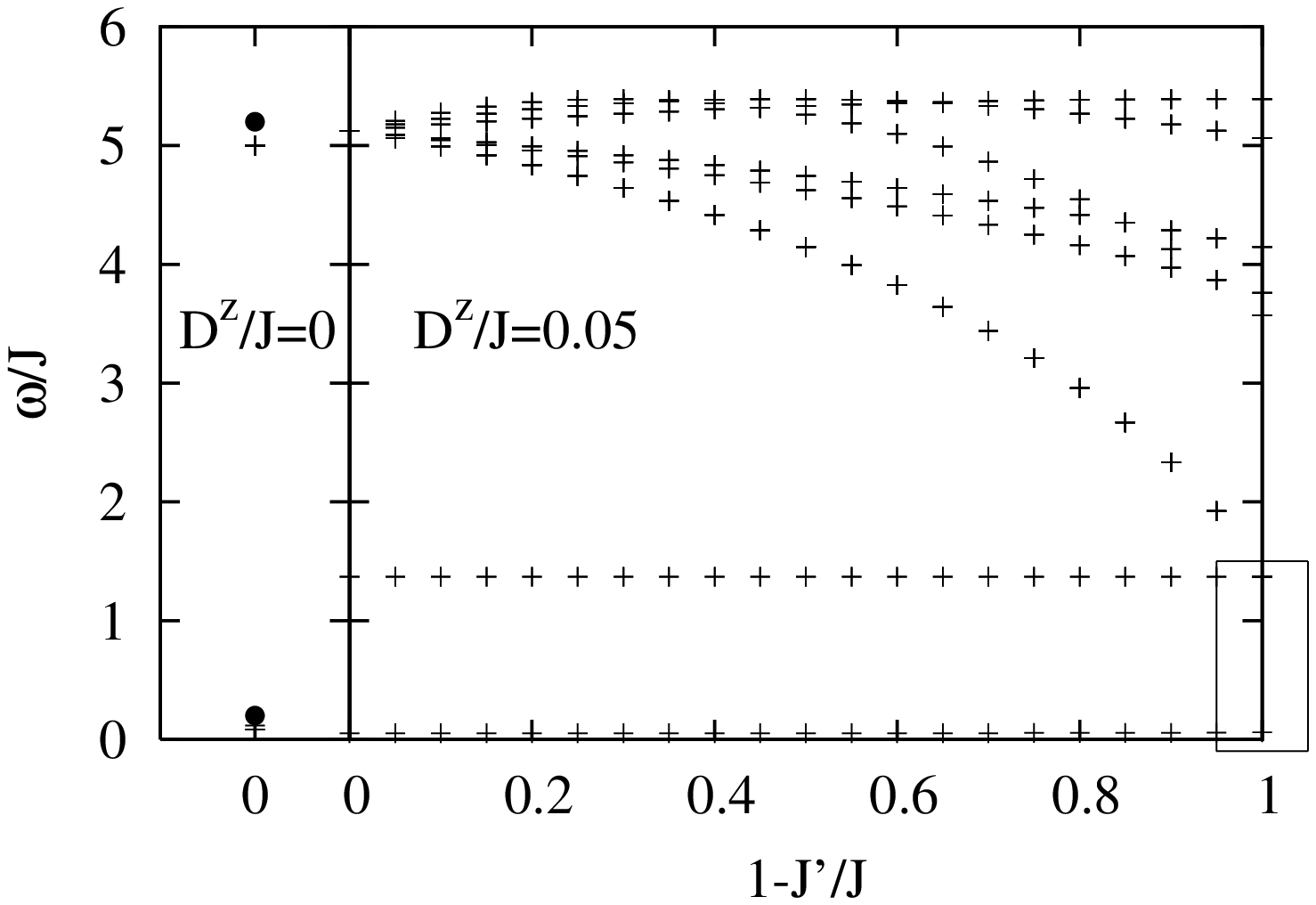}
 \includegraphics[width=5.5cm,angle=-90]{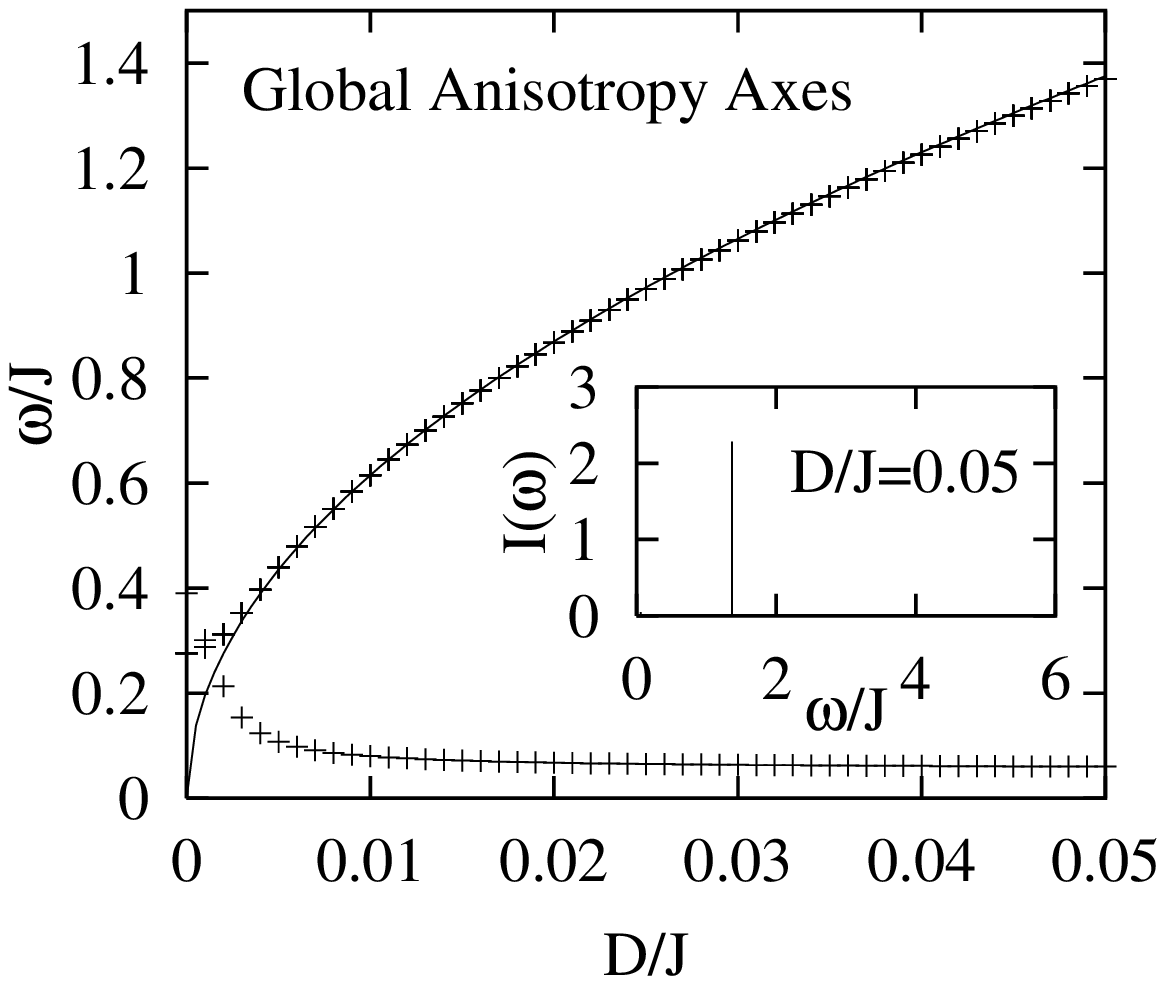}}
\caption{Left: Low-energy modes vs. model parameter $J^{\prime}/J$,
where $J^{\prime}$ corresponds to coupling all the spins of a given
sublattice to all the spins of another sublattice, for
$D^z/J=0;0.05$. Comparison is given with the exact triplet energies of
the fully connected model (solid black circles, $J'=J$,
$D^z/J=0$). The spectrum of the fully connected model\cite{Schnack}
($J'=J$) is different from that of the nearest neighbor Heisenberg
model ($J'=0$). \hfill\newline Right: Two lowest energy modes
vs. anisotropy $D/J$ for the nearest-neighbor model ($J'=0$) (note the
scale: only the low frequency range of the left fig. is shown
[corresponding to the box]).  The second energy is very close to the
$\sqrt{D}$ behavior (solid line) of the classical antiferromagnetic
resonance.  Magnetic-dipole ESR intensities of the modes are given in
the inset.}
\label{modesJall}
\end{figure}

\textit{Excitation Spectrum and ESR intensities}. We start with the
fully-connected model introduced in Ref. \cite{Schnack} where all the
spins of a given sublattice are connected to all the spins of an other
sublattice, $\frac{J}{5} \left(
\tilde{\textbf{\mbox{S}}}_A.\tilde{\textbf{\mbox{S}}}_B +
\tilde{\textbf{\mbox{S}}}_B.\tilde{\textbf{\mbox{S}}}_C+\tilde{\textbf{\mbox{S}}}_C.\tilde{\textbf{\mbox{S}}}_A
\right)$, where $\tilde{\textbf{\mbox{S}}}_A$ is a super-spin obtained
by adding $N/3$ spins $S$ on sublattice $A$. The Hamiltonian
factorizes and the energies are simply extracted from $ \frac{J}{10}
\left(\textbf{\mbox{S}}^2 - (S_A^2 + S_B^2 +S_C^2) \right)$
($\textbf{\mbox{S}}=\tilde{\textbf{\mbox{S}}}_A+\tilde{\textbf{\mbox{S}}}_B+\tilde{\textbf{\mbox{S}}}_C$). The
excited states form separated rotational bands with energies depending
upon $S(S+1)$,\cite{Schnack,Waldmann1} the lowest of which is
constructed by combining the three $S_{A,B,C}=NS/3$ maximum
spins. These lowest states are precisely the tower of states that were
found in exact numerical diagonalization of the triangular lattice
with nearest neighbor interactions.\cite{Lhuillier} These states would
collapse onto the classical ground state if the size of the system $N$
were allowed to go to infinity.\cite{Lhuillier} The ground state
energy per spin is exactly $-6.5J$ and among all the states of the
rotational bands there are three triplet states at $0.2J$ and six at
$5.2J$.\cite{Schnack} By the approximate method of section
\ref{method}, we have obtained a ground state energy of $-6.498J$ and
three states at $0.11J$ and $0.08J$ (twice degenerate)\footnote{The
degeneracy of the low-lying triplets is lifted with respect to the
exact solution, because discarding the quartic terms in the operators
$a$, $a^{\dagger}$ in the Hamiltonian has broken the total spin
symmetry.}  and the others at $5.0J$, which is in overall good
agreement (Fig. \ref{modesJall}, left, $D^z/J=0$).

We now consider the more realistic model (\ref{energy}) where the
additional couplings $J^{\prime}$ introduced in the previous paragraph
are reduced to zero. We see that the degenerate excited states at
$5.0J$ are split when one reduces $J^{\prime}$ (Fig. \ref{modesJall},
left). The final spectrum at $J^{\prime}=0$ is very different from
that of $J^{\prime}=J$ and the new gaps have nothing to do with the
original gap, $5.0J$. To explain the occurrence of an ESR line, we now
consider the \textit{global} single-ion anisotropy. The spectrum then
acquires a resonance frequency that scales as $\sqrt{DJ}$ for $D/J$
larger than about $0.003$ (Fig. \ref{modesJall}, right).  The
prefactor is numerically close to $S^2$, $\omega = (2.48)^2
\sqrt{DJ}$. This is similar to the \textit{antiferromagnetic resonance}
of the ordered antiferromagnets with single-ion
anisotropy. For instance, for the infinite triangular
lattice, we have $\omega = 3\sqrt{2DJ}$. When we suppress the Lagrange
multipliers in the present system (which means that we restore the
symmetry breaking of the N\'eel state), the frequency of the mode is
almost unchanged provided $D/J \gtrsim 0.003$.  Note that the Lagrange
multipliers are still important in the present context to find the
tower of states and at very small $D$; but if one wants to calculate
only the {\it frequency} of the antiferromagnetic resonance, one can
consider the broken-symmetry state as being a good approximation.
Nonetheless, to calculate the ESR intensity of a magnetic-dipole process,
\begin{equation}
I^{\alpha}(\omega) = \sum_e |\langle 0 | \sum_i S^{\alpha}_i | e \rangle|^2 \delta (\omega -\omega_e ),
\end{equation}
it is necessary to calculate not only the eigenvalues but also all the
eigenvectors. That forces us to introduce proper Lagrange multipliers
to avoid the singular transformation by suppressing the Goldstone
modes. This calculation shows that the
intensity is mainly in the \textit{antiferromagnetic resonance} (and
not in the other modes, in particular not in the higher energy modes)
as shown in the inset of Fig. \ref{modesJall}, right. The lowest
energy mode also has an intensity, but it is much smaller. At this stage,
it seems that everything is consistent with the recent ESR experiments
where a single peak has been observed.\cite{NojiriPrivate} This may be
an indication that the ground state of the system has indeed strong
\textit{coplanar} 120$^{\circ}$ short-range correlations as
we have shown.  We will, however, consider now a more realistic model
that leads to a more complicated excitation spectrum with additional
peaks reflecting the \textit{tangential} 120$^{\circ}$ correlations.

\subsection{Model with Local Anisotropy Axes}
\label{localanisotropy}

\begin{figure}[htbp]
 \includegraphics[width=6cm,angle=-90]{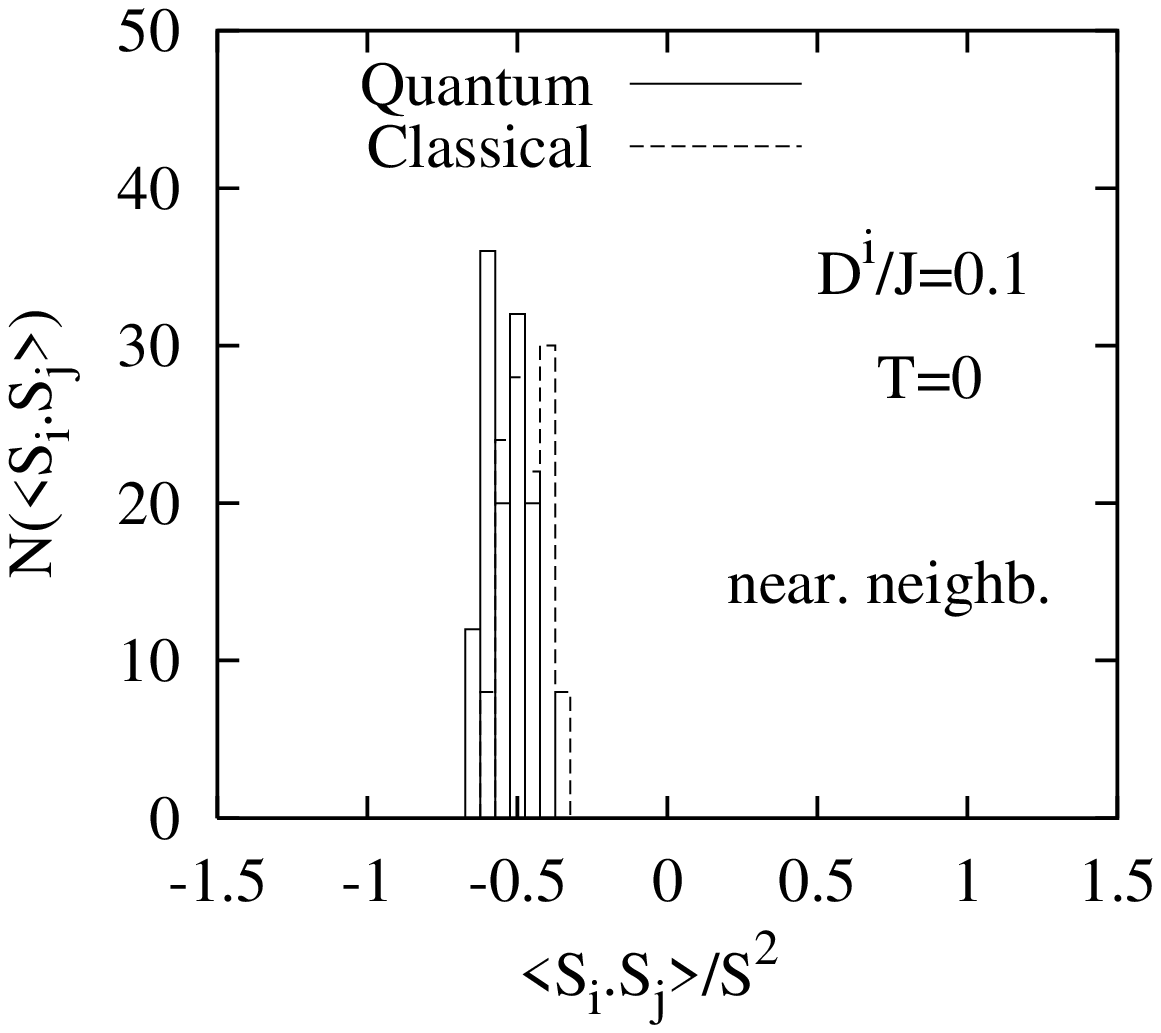} 
 \includegraphics[width=6cm,angle=-90]{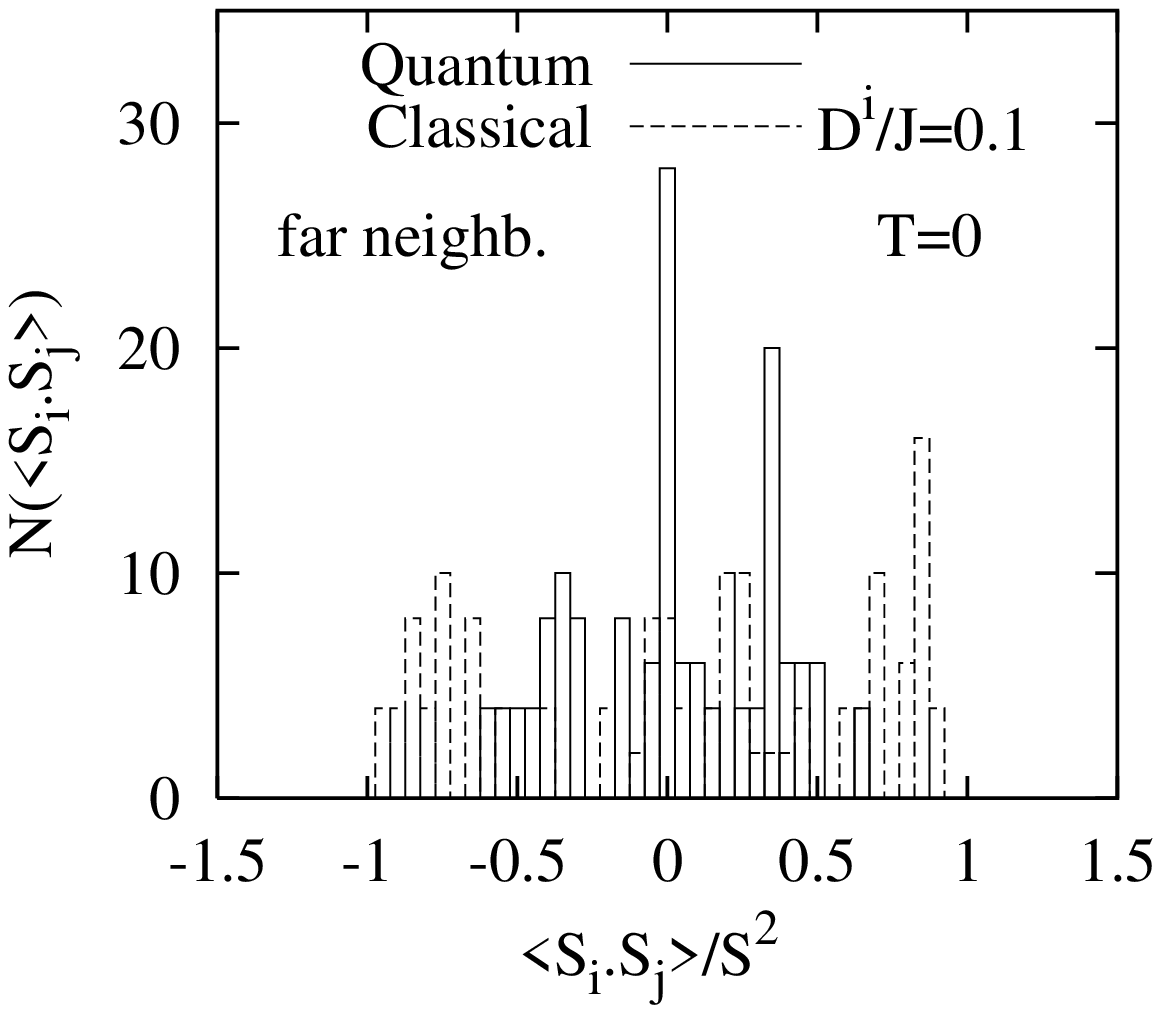}
\caption{Number of bonds with a given value of the correlation
function for the nearest neighbors (left) and the furthest-away bonds
(right). Local anisotropy axis. For definition of ``quantum'' and ``classical'', see fig. caption \ref{correlationsua}.}
\label{Psisjla}
\end{figure}

We now consider a more realistic model where the vectors associated
with D vary from site to site (model 2).  We argue that the simple
spectrum found in the previous section becomes more complicated and
many ESR lines should be visible. Starting now from the mean-field
ground state of Fig. \ref{spins} right, we calculate the correlations
when the quantum fluctuations are added. In Fig. \ref{Psisjla}, we
show the distribution of scalar products $\langle
\textbf{\mbox{S}}_i. \textbf{\mbox{S}}_j \rangle$.  For the nearest
neighbors, they are almost identical to the classical scalar products,
thus confirming the very strong correlations between the nearest
neighbors (Fig. \ref{Psisjla}, left). For neighbors that belong to
opposite sides of the sphere, the correlations have been reduced and
the distribution has a large peak at zero, but is still broad
(Fig. \ref{Psisjla}, right). For this model, correlations between far
neighbors are weaker.

\begin{figure}[htbp]
\centerline{
\includegraphics[width=6cm,angle=-90]{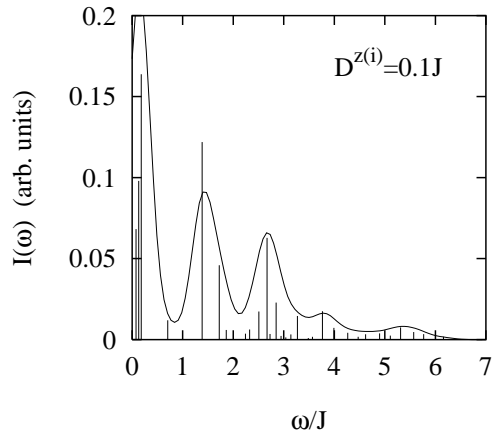}}
\caption{ESR intensity of the modes of energy $\omega$ (spikes) for
the model with \textit{local} anisotropies. The curve is a convolution
of the spikes with Gaussian functions.}
\label{intensity}
\end{figure}

Concerning the thermodynamic quantities, the specific heat, for
instance, is very similar to that of the simpler model with global
anisotropy, and weakly dependent upon the strength of the anisotropy
(Fig. \ref{specific-heat}). In particular we note that it would not be
possible to distinguish between the different models on the basis of
the specific heat only. However, the ESR spectrum shows different
features. We have calculated the frequencies and their intensities
(Fig. \ref{intensity}). All the frequencies get some intensity.  There
is indeed no selection rule such as $\Delta q =0$ in the present case
where all sites have a different classical magnetization. This is in contrast to
the classical 3-sublattice state previously discussed  where the
correlations
are simpler. Experimentally it is  plausible that the states are
mixed and appear as a broader line (see the convolution with Gaussian
functions in Fig.\ref{intensity}).

\section{Conclusion}

We have addressed the role of the quantum fluctuations and the
anisotropy in a complex molecular compound, Mo$_{72}$Fe$_{30}$.
First, we note that the specific heat at low temperatures is strongly
suppressed by quantum effects and weakly dependent upon the strength
of the anisotropy, contrary to classical approaches.\cite{Hasegawa}

The ground state depends, however, upon the anisotropy we choose.
We have described the short-range correlations in the ground state and
shown that the correlation length at zero-temperature is of the order
of the size of the molecule.  A local anisotropy which respects the
geometry of the sphere favors \textit{tangential} short-range
correlations (as in Fig. \ref{spins}, right), whereas global
anisotropy axes or, possibly, fluctuations would favor
\textit{coplanar} correlations (Fig. \ref{spins}, left). The nature of
the correlations could be directly tested by elastic neutron scattering.

Moreover, we have shown that the excitation spectrum exhibits
different features depending on the nature of the ground state.  When
the spins are coplanar, the spectrum consists of
an \textit{antiferromagnetic resonance} (as in classical systems) with
a $\sqrt{DJ}$ behavior for $D$ not too small, separated from higher
magnon states by a sizeable gap. Magnetic-dipole ESR transitions are
found to be allowed from the ground state to the
antiferromagnetic resonance  only. The other states should
be visible in inelastic neutron scattering, for instance.

For a ground state with tangential correlations, the excitation
spectrum is more complex and a quasi continuum of magnon states is
found at low energy (fig. \ref{intensity}). In this case, because all
the sites have different spin directions (and the symmetries relating
different sites are completely broken), there is no selection rule for
magnetic-dipole transitions: all the modes acquire some intensity.
Combined with other relaxational mechanisms, the overall effect would
be to give a very broad ESR signal, as indicated in the
figure. Unusually large broadening seems to be observed
experimentally\cite{NojiriPrivate}. It could be taken as an indication
that the ground state has \textit{tangential} short-range correlations
rather than \textit{coplanar}, although, strictly speaking, the
comparison needs to be done at finite fields.

A general feature of finite-size systems is the tower of states at
very low energy. Takahashi's method provides a way to go beyond the
scaling limit result for calculating the energy of the magnon modes of
the tower of states, and we have given their energy. It would be
interesting to observe these modes, which would explain the slow
dynamics of these systems.

\section*{Acknowledgments}

It is a pleasure to thank Prof. H. Nojiri who has stimulated this work and
who has provided us with his experimental data prior to publication.

\end{document}